\documentclass[cameraready]{Interspeech}

\title{Assessing the Impact of Noise and Speech Enhancement on the Intelligibility of Speech Codecs}

\author[affiliation={1}, orcid=0009-0005-2519-3579, 
]{Lyonel}{Behringer}
\author[affiliation={1}]{Anna}{Leschanowsky}
\author[affiliation={1}]{Anjana}{Rajasekhar}
\author[affiliation={1}]{Emily}{Kratsch}
\author[affiliation={1}, orcid=0009-0009-1045-6064]{Guillaume}{Fuchs}

\address{
    $^1$ Fraunhofer Institute for Integrated Circuits (IIS), Erlangen, Germany
}

\email{lyonel.behringer@iis.fraunhofer.de
}


\newcommand{\red}[1]{\textcolor{red}{#1}}
\usepackage{comment}
\usepackage{graphicx}
\usepackage{cite}
\usepackage{balance}

\begin{document}

\maketitle

\begin{abstract}
Preserving speech intelligibility is a minimum requirement for speech codecs in communication. Recently, very low-bitrate neural codecs have gained interest for replacing classical codecs, reinforcing the need to evaluate whether intelligibility is preserved in realistic scenarios.
In this paper, we evaluate the intelligibility and listening effort of classical and neural speech codecs in clean and noisy conditions. Further, we assess the impact of speech enhancement (SE) before coding, simulating a possible audio processing pipeline.
The results show that classical codecs are more noise robust than neural codecs. Further, SE can lead to significant intelligibility and listening effort improvements for codecs otherwise negatively affected by noise. Listening effort reveals nuanced differences when intelligibility is saturated. 
Lastly, objective intelligibility based on automatic speech recognition is highly correlated with subjective intelligibility scores averaged per condition.

\end{abstract}

\section{Introduction}

Neural speech codecs (NSCs) have recently gained popularity due to their capability of coding speech at lower bitrates than classical codecs.
Oftentimes, proposed NSCs have been evaluated only in clean speech conditions and without distinguishing between performance in clean and noisy conditions~\cite{valin2019lpcnet, kumar23_dac, defossez2024_moshi, liu24_semanticodec, parker2025_stablecodec, wu25f_ts3codec}.
Moreover, the assessment of overall speech quality using either subjective methods such as ~\cite{ITU-BS1534-3-2015, ITUTP800} or objective methods such as ~\cite{chinen2020visqol, ragano2024scoreq} is most prevalent in NSC evaluation, although assessing speech intelligibility is also recommended \cite{schmidt1995intelligibility, lechler2024_icassp_crowd_si} and corresponding toolkits are publicly available~\cite{leschanowsky25_sitool}. For generative NSCs, this is relevant already for clean speech due to potential content hallucinations, which may not be reflected in speech quality, especially in no-reference tests. Intelligibility assessment becomes even more important in more challenging scenarios relevant in real-time communication, e.g. low-delay constraints, noisy environments, with or without the use of speech enhancement (SE).
In such cases, explicit intelligibility evaluation is warranted to guarantee seamless communication.

While gaining some traction recently, the research body on intelligibility of NSCs in noisy conditions (as well as for quality) remains scarce. \cite{zheng2025_quant_perturb, tseng25_probing_robustness} rely on purely objective metrics for evaluating intelligibility in noise, namely STOI \cite{TaalSTOI} and word error rate (WER) of an automatic speech recognition (ASR) system, respectively. In \cite{wojcicki2025_lrac}, subjective methodologies were used for evaluating the quality and word-level intelligibility of submitted NSCs in clean and noisy conditions.
To the best of our knowledge, there is no work subjectively evaluating the intelligibility of NSCs at the sentence level, i.e. reflecting a real listening situation with contextual information.

In domains other than speech coding, various works have conducted sentence-level intelligibility assessments in adverse listening conditions,
such as SE~\cite{SI_of_SE_algorithms, 2023_CEC2}, intelligibility prediction~\cite{barker25_clarity}, or near-end listening enhancement~\cite{rennies2018_nele_SI_LE}. In contrast to word-level tests, which can offer phoneme-specific insights~\cite{ANSIASAS3.2:2020_drt_mrt_pbwt}, sentence-level tests using naturalistic sentences have the advantage of representing real-world communication scenarios~\cite{billings2023speech_in_noise_testing}, which are a common use case for speech codecs. They constitute open-response sets, which is a factor increasing test difficulty compared to closed-response sets. On the other hand, contextual cues can reduce the difficulty~\cite{nilsson1994development_hint}. A commonly mentioned challenge of evaluating open-response sets is the scoring of transcripts, e.g. regarding spelling differences, which might vary depending on the conducted research~\cite{baese2023_intelligibility}. 
Moreover, the same sentence should not be repeated within the same session to avoid learning effects~\cite{nilsson1994development_hint}.

Across methodologies, ceiling effects are a well-known difficulty of intelligibility assessment \cite{schmidt1995intelligibility, leschanowsky25_sitool, rennies2018_nele_SI_LE}. Recommended remedies include the use of open-response sets \cite{schmidt1995intelligibility} as well as the assessment of additional information related to speech processing such as listening effort \cite{rennies2018_nele_SI_LE, baese2023_intelligibility}. We apply these recommendations in this work.

Further, since subjective intelligibility evaluation is costly and time-consuming, reliable objective metrics are desirable. \cite{leschanowsky25_sitool} compared subjective and objective intelligibility results of NSCs in clean conditions, showing good correlations for STOI and ESTOI~\cite{jensen2016_estoi}, but not for WER of ASR systems. No such comparison has been made for NSCs in noisy conditions. In addition, single-channel SE of noisy speech can negatively impact the WER of ASR systems ~\cite{ochiai2024_impact_SE_WER}, but we are not aware of any comparisons to subjective evaluations.


The contributions of this work are: 1) We perform a systematic crowdsourced evaluation of intelligibility of diverse neural and classical speech codecs at the sentence level, across multiple noise categories and signal-to-noise ratios (SNR).
2) We assess listening effort to investigate its usefulness for resolving intelligibility ceiling effects.
3) We assess the impact of SE preprocessing on intelligibility and listening effort. 4) Finally, we correlate the subjective results with a range of objective metrics.

\section{Experiments}
\subsection{Benchmarked Codecs}
\label{sec:BenchmarkedCodecs}

\begin{table}[h!]
\caption{Codecs under test. CPU usage is measured on a single core of an Intel Core i7‑11700 at \SI{2.5}{kHz} using publicly available implementations. Number of parameters is given for neural codecs when available.}
\label{tab:benchmarked_codecs}
\centering
\resizebox{\columnwidth}{!}{%
\begin{tabular}{l c c c c}
\hline
\textbf{Codec} & \textbf{kbps} & \textbf{\#Params} & \textbf{CPU usage} & \textbf{Alg. Delay} \\
\hline
AMR-WB    & 6.6 & - & 0.87\% & \SI{26}{ms} \\
EVS       & 8   & - & 1.50\% & \SI{32}{ms} \\
\hline
LPCNet    & 1.6 & 71.6K & 23.87\% & \SI{65}{ms} \\
Lyra V2    & 3.2 & - & 11.25\% & \SI{20}{ms} \\
DAC       & 1.5 & 76M & \red{210.63\%} & \red{non-causal} \\
Mimi      & 1.1 & 80M & \red{147.50\%} & $\geq$ \SI{80}{ms} \\
\hline
\end{tabular}}

\end{table}

Table~\ref{tab:benchmarked_codecs} lists all codecs under test. We selected conventional speech codecs representative of current real‑world communication systems, including two generations of 3GPP codecs: \textbf{AMR‑WB}~\cite{amr-wb} and \textbf{EVS}~\cite{evs}, evaluated at \SI{6.6}{kbps} and \SI{8}{kbps}, respectively, corresponding to their minimum or near‑minimum operating bitrates. Both rely on the CELP paradigm, a hybrid approach that aims to preserve the input waveform through analysis-by-synthesis optimization of the coded parameters.

For very low bitrate neural codecs, we selected open-source codecs available at the time of the study and representing various levels of complexity, some of which could also be used in real-time communication as their architecture is causal and processing streamable, while their complexity is low enough for real-time processing on mobile devices.

First, we consider \textbf{LPCNet}\footnote{\url{https://github.com/xiph/LPCNet/commit/7dc9942}}~\cite{valin2019lpcnet}, an early neural coding solution combining classical signal processing and deep neural networks to decode a bitstream generated by a conventional encoder, operating at \SI{1.6}{kbps}. Its relatively high complexity lies in its autoregressive approach and sample-by-sample generation, but it can nevertheless operate in real time on a CPU.
We also evaluate various GAN-based end-to-end autoencoder approaches, a paradigm introduced by SoundStream~\cite{zeghidour2022_soundstream} built upon a Residual Vector Quantization (RVQ) of the latent. This principle was subsequently adopted by most state-of-the-art neural speech codecs. \textbf{Lyra V2}\footnote{\url{https://github.com/google/lyra/tree/v1.3.2}}~\cite{lyra}, an open-source codec derived from SoundStream, is optimized to run in real time on a smartphone CPU. It is tested at \SI{3.2}{kbps} and is the most computationally efficient neural codec under test. 
We also consider \textbf{DAC}~\cite{kumar23_dac}, a much more complex and non-causal variant of the same paradigm, which prevents its use for real-time communication applications, but demonstrates better quality. We selected the speech fine-tuned version of DAC proposed in ~\cite{shechtman24_dac_ibm} and trained at \SI{1.5}{kbps}. In clean conditions, it represents one of the best qualities achievable at this bitrate.
The final neural codec under test is \textbf{Mimi}~\cite{defossez2024_moshi}, evaluated at \SI{1.1}{kbps}.  It employs advances like a transformer-based architecture at the bottleneck and latent semantic disentanglement obtained through distillation.
Mimi is causal but has a computational complexity similar to that of DAC, which makes it unsuitable for mobile devices.



\subsection{Test Procedure} 

\subsubsection{Selection and Preparation of Test Items}

We use a subset of the Clarity Speech Corpus~\cite{Graetzer2022_clarity} (CSC) for the crowdsourced test, as it offers naturalistic sentences and is established in the hearing aid domain~\cite{2023_CEC2, barker25_clarity}.

The test items before coding are comprised of clean, noisy, and enhanced noisy speech. In detail, the test material consists of 12 unique sentences from 4 randomly selected speakers (2 male, 2 female), respectively, i.e. 48 unique sentences overall. Phonemic balance per speaker is approximated by selecting sentences via the mLTM algorithm~\cite{suyanto2006_mltm} to ensure adequate coverage of phonemes. The speech items are padded with \SI{2}{\second} leading and trailing silence. All files are loudness normalized to \SI{-24}{dBov} using sv56~\cite{ITU_G191_2005}. For noise, we use four representative noise types from the DEMAND database~\cite{thiemann_demand}: DLIVING (living room), PRESTO (restaurant babble), TCAR (car engine), and TMETRO (metro). 
Each speaker's clean items are mixed with the four noise types in equal proportions, resulting in 12 unique sentences per noise type.
Noise mixing is done with 5, 15, and \SI{25}{dB} SNR. 
Noisy items are included without SE as well as with SE\footnote{We did not enhance clean items as informal comparisons showed virtually no difference.} via DeepFilterNet2~\cite{schröter2022deepfilternet2}, a real-time capable SE model with a complexity of around \SI{0.7}{GFLOPs}. All items are then resampled to \SI{
16}{\kilo\hertz} sampling rate and processed by all codecs\footnote{Since DAC and Mimi operate at \SI{24}{kHz}, an intermediate resampling stage is used for them.}, resulting in a total of 2,352 test items. 

\subsubsection{Test Configuration and Test Interface}

Test sessions were created using an incomplete block design approach. For each session, 48 test items were randomly selected, ensuring that participants heard each unique sentence only once to avoid learning effects. In addition, each session included three trap questions. Participants who answered at least one trap question incorrectly were excluded from the analysis.

A modification of SITool~\cite{leschanowsky25_sitool} was used to collect responses of the participants. The application presented audio stimuli and allowed participants to transcribe the heard sentence. To resolve potential ceiling effects, participants were also asked to rate listening effort on a five-point scale according to the ITU-T Recommendation P.800 Annex B~\cite{ITUTP800} (1 = "no meaning understood with any feasible effort"; 5 = "complete relaxation possible; no effort required"). 
Replay of audio samples was allowed.
Participants were instructed to enter ``not understood'' if they were unable to understand any part of the audio.


\subsection{Transcript Normalization and Score Calculation}
All transcripts are normalized by lowercasing, removing punctuation and excess whitespaces, and applying number-to-grapheme conversion. 

As is common for sentence-level intelligibility, the speech intelligibility score (SI) is calculated as 
\begin{align}
SI = \frac{W_{c}}{W_{t}},
\end{align}
where ${W_{c}}$ is the number of correctly transcribed words and ${W_{t}}$ is the number of reference words. 
Since SI does not account for insertions, the word error rate (WER)\footnote{\url{https://github.com/jitsi/jiwer}} is also computed.

\subsection{Participants and Screening Procedures}

The test was deployed on the crowdsourcing platform Amazon Mechanical Turk. Participation was restricted to workers located in native English-speaking countries. Workers received an hourly compensation of \$10.50 USD, aligning with prior studies~\cite{lechler2024_icassp_crowd_si}. Informed consent was obtained from all participants before the test, along with demographic information on age and gender. To ensure adequate English language proficiency, a preliminary screening was included at the beginning of the session. This consisted of four clean, unprocessed CSC sentences not used in the main test. Only participants achieving a WER of 10\% or lower on the screening task were allowed to proceed to the main test. 65 participants were excluded in this manner.
To minimize low-effort responses such as incomplete or nonsense transcriptions in the main test, a post-screening was also applied. Participants who did not achieve a WER of 30\% or lower for the clean part of the main test were excluded (7), as well as  nonsensical random-letter transcriptions (10). The test was open until a minimum of three valid responses were obtained for each item to allow for assessment of inter-annotator reliability (IAR). The clean and \SI{25}{dB} SNR reference stimuli have acceptable IAR ($\alpha = 0.67 - 0.75$). For the reference at lower SNRs, IAR decreases, suggesting an increase in task difficulty as expected. However, as we have incomplete listener overlap, listener variability cannot be ruled out as a contributing factor. 
In total, 7,670 valid responses by 160 participants were collected.

\subsection{Correlation with Objective Metrics}
We consider several objective metrics for correlation analysis. As established intrusive metrics for intelligibility, we use STOI and ESTOI, where we always compare test items to the respective clean reference items. Further, we use ASR transcripts to compute objective SI. The evaluated ASR models span complexities from 74M to 1.5B parameters, namely Whisper~\cite{radford2023robust} base (Whisper-B) and large-v3 (Whisper-L), as well as Parakeet-TDT-0.6B-v3 and Canary-1B-v2~\cite{sekoyan2025_canary1bv2_parakeettdt06bv3}.

We apply a 3rd-order monotonic polynomial mapping following ITU-T P.1401~\cite{ITU-T_P.1401} and compute the Pearson correlation coefficient (PC), Spearman's rank correlation coefficient (SC) and root mean square error (RMSE) between subjective and objective results, both for individual samples (sample-wise) as well as for results aggregated by codec-SE-SNR combination (condition-wise).

\section{Results}
\label{sec:Results}

\subsection{Subjective Speech Intelligibility Evaluation}
\label{sec:Results_subj_SI}

Figure~\ref{fig:subj_SI_LE_overall} depicts the subjective intelligibility results. As subjective SI and WER scores showed a PC of $0.99$, WER is omitted. A linear mixed-effects model (LMM) was fitted to statistically analyze the effects of codec, noise type, SNR level, and SE on intelligibility. These factors were included as fixed effects, along with a codec-by-SE interaction term, as the impact of SE was expected to vary depending on the codec used. A random intercept for sentence ID was included to account for variability across sentences. Holm correction was applied when testing multiple pairwise contrasts.

\subsubsection{Impact of Codec on Intelligibility}

As depicted in Figure~\ref{fig:subj_SI_LE_overall}, intelligibility scores are near ceiling for the reference and all codecs in the clean condition and at \SI{25}{dB} SNR. We perform pairwise contrasts to EVS without SE as it is the overall best-performing codec. Significant ($p < 0.05$) differences between codecs are found at 15 and \SI{5}{dB} SNR. At \SI{15}{dB} SNR, EVS scores significantly higher than LPCNet without SE and Mimi without SE. At \SI{5}{dB} SNR, EVS is significantly better than DAC without SE, LPCNet with/without SE, and Mimi with/without SE. Pairwise contrasts to the reference and LPCNet without SE at \SI{5}{dB} SNR show that the reference is not significantly different from the classical codecs but from all neural codecs without SE, and that LPCNet is reliably outperformed by every other codec. These results in conjunction with Figure~\ref{fig:subj_SI_LE_overall} demonstrate that classical codecs are more noise robust than neural codecs, with the difference becoming larger as SNR decreases. 




\subsubsection{Impact of SE on Intelligibility}

\begin{figure}
    \centering
    \includegraphics[width=\columnwidth]{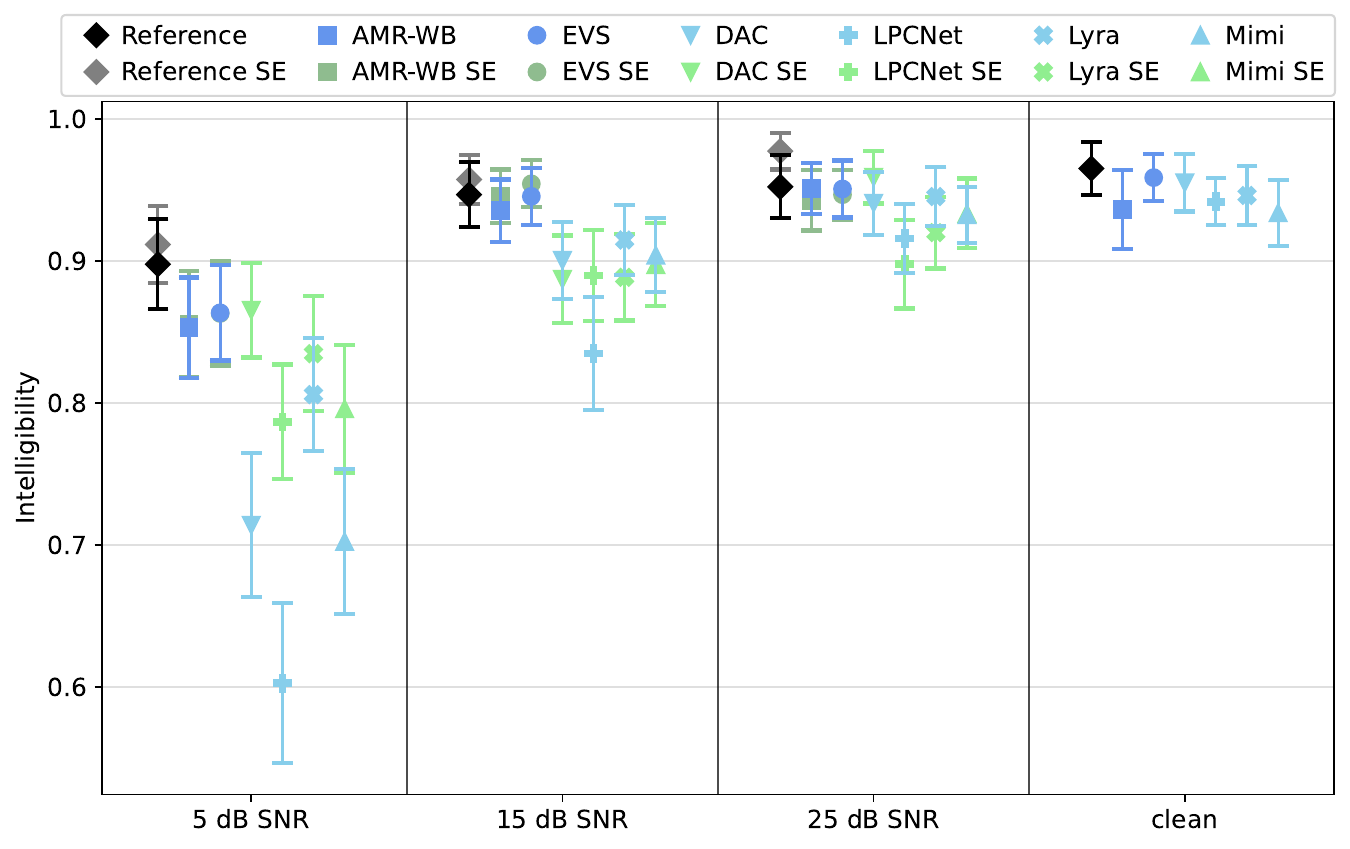}
    \caption{Subjective mean intelligibility score for clean and noisy conditions with confidence intervals, with and without SE.}
    \label{fig:subj_SI_LE_overall}
\end{figure}

The significance of the SE effect on each codec was assessed by computing the difference between the results, with and without SE, predicted by the LMM considering  SE main effect and the codec-by-SE interaction term.
Significant intelligibility improvements were observed for DAC ($\Delta = .060$, $z = 4.93$, $p < .001$), 
LPCNet ($\Delta = .082$, $z = 8.78$, $p < .001$), 
and Mimi ($\Delta = .036$, $z = 2.93$, $p = .003$). 
The effects for AMR-WB, EVS, Lyra, and the reference were non-significant.
As illustrated in Figure~\ref{fig:subj_SI_LE_overall}, SE improvements are largest at lower SNRs where the intelligibility of neural codecs is most deteriorated. These results demonstrate that while classical codecs are more noise robust than neural codecs, SE preprocessing can reduce this gap, enabling high intelligibility in adverse conditions at reduced bitrates. 


\subsubsection{Impact of Noise Type on Intelligibility}

\begin{figure*}
    \centering
    \includegraphics[width=0.84\linewidth]{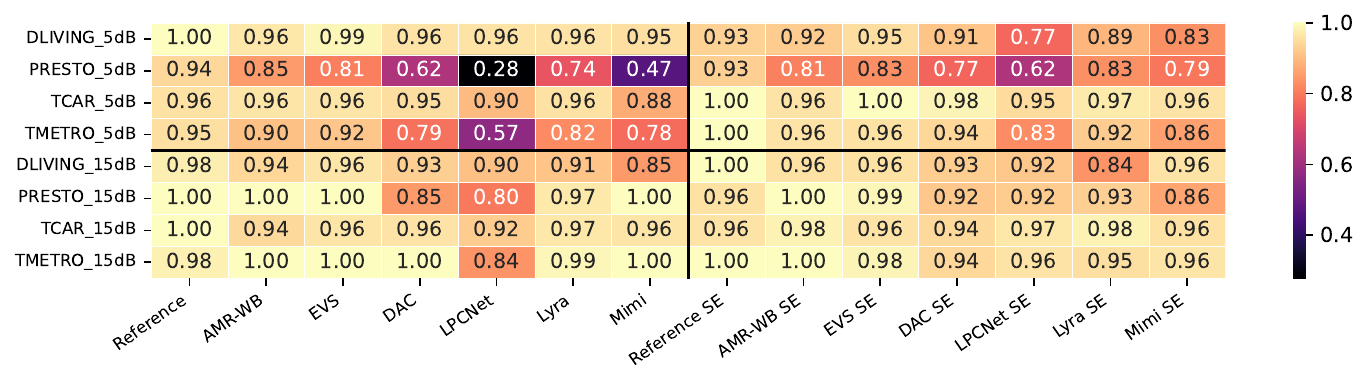}
    \caption{Noise-specific median SI. The x-axis indicates the codec and potential SE, the y-axis noise type and SNR.}
    \label{fig:heatmap_ssi}
\end{figure*}

Figure~\ref{fig:heatmap_ssi} shows a heatmap of noise-specific SI scores. Due to the limited amount of items per noise type, we report median values and highlight only substantial differences.
The results for \SI{25}{dB} SNR exhibit no noise-specific differences and are omitted. 

The two most detrimental noise types are PRESTO and TMETRO, which  are both very rich in frequency coverage. At \SI{15}{dB} SNR without SE, LPCNet shows SI deterioration for these noises compared to the reference, while DAC shows degradation for PRESTO. At \SI{5}{dB} SNR, the impact becomes more severe for the neural codecs. EVS also shows degradation for PRESTO, albeit less than for the neural codecs. TCAR and DLIVING 
are the least problematic noise types. TCAR is concentrated to the frequencies between 0 and \SI{100}{\hertz}, resulting in very little overlap with speech, while DLIVING contains  living room noises and background music. 

SE improves the SI of DAC, LPCNet, and Mimi for PRESTO, as well as DAC and LPCNet for TMETRO. Conversely, SE deteriorates the SI of LPCNet for DLIVING at \SI{5}{dB} SNR. 
In summary, the codecs have varying robustness to different noises, and SE predominantly maintains or improves SI.
    
\subsection{Subjective Listening Effort Evaluation}

An LMM as in Section ~\ref{sec:Results_subj_SI} was fitted for listening effort. 

\subsubsection{Impact of Codec and SE on Listening Effort}
Regarding the impact of codec across SNRs, the results for listening effort parallel those for intelligibility.
Based on the model-estimated SE and codec-by-SE interaction effects, SE significantly improved listening effort for LPCNet ($\Delta = .490$, $z = 7.65$, $p < .001$), DAC ($\Delta = .284$, $z = 4.44$, $p < .001$), Mimi ($\Delta = .175$, $z = 2.72$, $p = .006$), and EVS ($\Delta = .169$, $z = 2.63$, $p = .009$). 
 SE effects were not reliable for AMR-WB, Lyra, or the reference. Such similar results are expected due to the relatedness of intelligibility and listening effort.

\subsubsection{Resolving Intelligibility Ceiling Effects}
\begin{figure}
    \centering
    \includegraphics[width=\columnwidth]{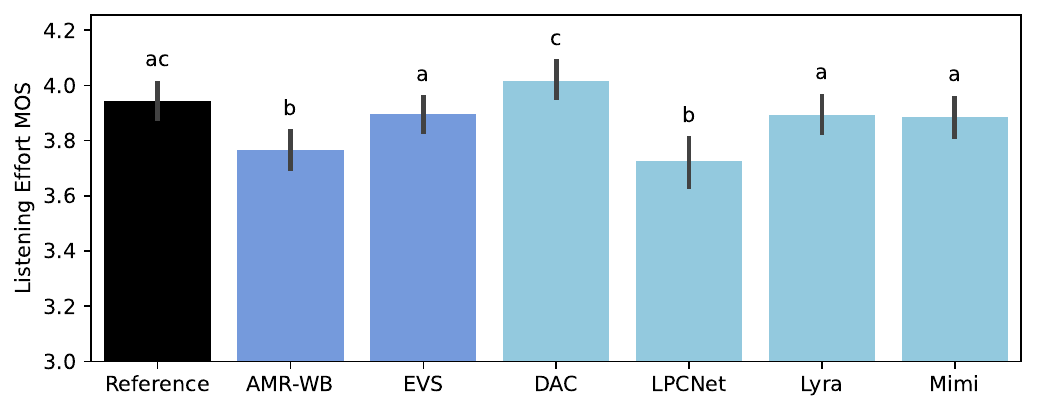}
    \caption{Listening effort MOS with compact letter display for subset where $SI >= .95$.  Higher MOS means less effort required.}
    \label{fig:LE_ceiling_effects}
\end{figure}

To assess whether listening effort can resolve SI ceiling effects and reveal nuances of listening experience, we fitted the LMM to a subset where $SI >= .95$.
Figure~\ref{fig:LE_ceiling_effects} illustrates the codec differences in terms of listening effort mean opinion score (MOS) with a compact letter display, i.e. different letters between two codecs indicate a significant ($p<0.05$) pairwise difference. While intelligibility shows no significant differences for this subset, DAC requires significantly less listening effort than all other conditions except the reference. AMR-WB and LPCNet require similar listening effort but significantly more than all other conditions. While listening effort constitutes a different dimension of listening experience than intelligibility, these results confirm the usefulness of listening effort as supplementary evaluation to intelligibility when facing ceiling effects.

\subsection{Correlation of Intelligibility with Objective Metrics}

For the correlation analysis, subjective results were averaged across listeners per item. Table~\ref{tab:subj_obj_corr} shows the PC, SC, and RMSE between subjective SI and objective metrics. For both sample-wise and condition-wise, subjective SI shows higher PC with ASR-based objective SI than with STOI and ESTOI. 
Whisper-B yields the best condition-wise PC and SC, as well as the highest sample-wise SC, while Whisper-L shows the best sample-wise PC and second-best SC and RMSE.

Overall, condition-wise PC, SC, and RMSE are substantially better than sample-wise. The results demonstrate that the evaluated objective metrics can be an efficient means of assessing sentence-based intelligibility of clean and noisy speech at condition level, whereas a sample-wise use for replacing subjective tests is questionable.
Whisper-B is particularly notable, as it is by far the least complex ASR model. This indicates that the use of less complex ASR models can be similarly or even better suited as a proxy for subjective intelligibility.


\begin{table}[h!]

    \centering
    \caption{Condition-wise (c.) and sample-wise (s.) PC, SC, and RMSE between subjective SI and objective metrics ($p < 0.001$). OSI = objective SI. Best scores in bold, second-best underlined.}
    \label{tab:subj_obj_corr}
\resizebox{\columnwidth}{!}{%
\begin{tabular}{lcccccc}
\toprule
\textbf{Objective metric} & \textbf{c.PC} & \textbf{c.SC} & \textbf{c.RMSE} & \textbf{s.PC} & \textbf{s.SC} & \textbf{s.RMSE} \\
\midrule
STOI & .870 & .891 & .039 & .445 & .364 & \textbf{.089} \\
ESTOI & .903 & .897 & .051 & .507 & .373 & .116 \\
OSI Whisper-B & \textbf{.973} & \textbf{.936} & .024 & .679 & \textbf{.519} & .152 \\
OSI Whisper-L & .941 & .881 & .025 & \textbf{.762} & \underline{.460} & \underline{.097} \\
OSI Canary & .946 & .854 & \underline{.021} & \underline{.704} & .405 & .101 \\
OSI Parakeet & \underline{.969} & \underline{.921} & \textbf{.017} & .702 & .430 & .112 \\

\bottomrule
\end{tabular}}
\end{table}

\section{Conclusion}
In this work, we conducted a crowdsourced evaluation of clean and noisy speech processed by multiple neural and classical speech codecs. We assessed speech intelligibility and listening effort and demonstrated that neural codecs are less noise robust than classical codecs.
Additionally, we showed that SE preprocessing of noisy speech benefits the intelligibility and listening effort of neural codecs which otherwise suffer from decreased performance, proving the effectiveness of such an audio processing pipeline. 
Given ceiling effects in intelligibility, listening effort was found to be a useful differentiating aspect of listening experience.
A noise-specific analysis revealed that codec robustness varies depending on the noise type.
Further, we analyzed the correlation of subjective intelligibility with multiple objective metrics and found that ASR systems are highly correlated with subjective intelligibility at condition level, outperforming the established STOI and ESTOI. 
A limitation of the test methodology is reduced inter-annotator reliability at very low SNR, which could be attributed to increased task difficulty and listener variability.
Future work could entail multilingual evaluation as well as dedicated model trainings to strictly assess how specific modifications such as training data or model architecture affect noise robustness regarding intelligibility. 

\newpage

\section{Generative AI Use Disclosure}
Generative AI was used for cosmetic improvements of Figures~\ref{fig:subj_SI_LE_overall} and~\ref{fig:heatmap_ssi}. Correctness of the plotting code was manually confirmed by the authors. 

\section{Acknowledgements}
\ifcameraready
     This research was partially supported by the Free State of Bavaria in the DSGenAI project  (Grant No.: RMF-SG20-3410-2-18-4). The authors thank Kishor Kayyar Lakshminarayana for his insightful feedback regarding test methodologies.
\else
     This section is anonymized for review.
\fi

\begingroup
\tiny
\bibliographystyle{IEEEtran}

\bibliography{mybib}

@STRING{icassp = {Proc. {IEEE} Intl. Conf. on Acoustics, Speech and Signal Processing
	(ICASSP)}}

@inproceedings{leschanowsky25_sitool,
  title     = {{Benchmarking Neural Speech Codec Intelligibility with SITool}},
  author    = {Anna Leschanowsky and Kishor {Kayyar Lakshminarayana} and Anjana Rajasekhar and Lyonel Behringer and Ibrahim Kilinc and Guillaume Fuchs and Emanuël A. P. Habets},
  year      = {2025},
  booktitle = {{Interspeech 2025}},
  pages     = {5488--5492},
  doi       = {10.21437/Interspeech.2025-984},
  issn      = {2958-1796},
}

@misc{zheng2025_quant_perturb,
      title={Enhancing Noise Robustness for Neural Speech Codecs through Resource-Efficient Progressive Quantization Perturbation Simulation}, 
      author={Rui-Chen Zheng and Yang Ai and Hui-Peng Du and Li-Rong Dai},
      year={2025},
      eprint={2509.19025},
      archivePrefix={arXiv},
      primaryClass={eess.AS},
      url={https://arxiv.org/abs/2509.19025}, 
}

@inproceedings{tseng25_probing_robustness,
  title     = {{Probing the Robustness Properties of Neural Speech Codecs}},
  author    = {Wei-Cheng Tseng and David Harwath},
  year      = {2025},
  booktitle = {{Interspeech 2025}},
  pages     = {5013--5017},
  doi       = {10.21437/Interspeech.2025-355},
  issn      = {2958-1796},
}

@misc{wojcicki2025_lrac,
      title={Low-Resource Audio Codec (LRAC): 2025 Challenge Description}, 
      author={Kamil Wojcicki and Yusuf Ziya Isik and Laura Lechler and Mansur Yesilbursa and Ivana Balić and Wolfgang Mack and Rafał Łaganowski and Guoqing Zhang and Yossi Adi and Minje Kim and Shinji Watanabe},
      year={2025},
      eprint={2510.23312},
      archivePrefix={arXiv},
      primaryClass={cs.SD},
      url={https://arxiv.org/abs/2510.23312}, 
}

@article{baese2023_intelligibility,
  title={Intelligibility as a measure of speech perception: Current approaches, challenges, and recommendations},
  author={Baese-Berk, Melissa M and Levi, Sarah V and Van Engen, Kristin J},
  journal={The Journal of the Acoustical Society of America},
  volume={153},
  number={1},
  pages={68},
  year={2023},
  month={January},
  doi={10.1121/10.0016806},
  pmid={36732227}
}

@article{rennies2018_nele_SI_LE,
  author={Jan Rennies and Arne Pusch and Henning Schepker and Simon Doclo},
  title={Evaluation of a near-end listening enhancement algorithm by combined speech intelligibility and listening effort measurements},
  journal={The Journal of the Acoustical Society of America},
  volume={144},
  number={4},
  pages={EL315--EL321},
  year={2018},
  publisher={AIP Publishing}
}

@inproceedings{lechler2024_icassp_crowd_si,
  author={Laura Lechler and Kamil Wojcicki},
  booktitle = {ICASSP 2024 - 2024 IEEE International Conference on Acoustics, Speech and Signal Processing (ICASSP)}, 
  title={Crowdsourced Multilingual Speech Intelligibility Testing}, 
  year={2024},
  pages={1441-1445},
  doi={10.1109/ICASSP48485.2024.10447869}
}

@INPROCEEDINGS{2023_CEC2,
  author={Akeroyd, Michael A. and Bailey, Will and Barker, Jon and Cox, Trevor J. and Culling, John F. and Graetzer, Simone and Naylor, Graham and Podwińska, Zuzanna and Tu, Zehai},
  booktitle={ICASSP 2023 - 2023 IEEE International Conference on Acoustics, Speech and Signal Processing (ICASSP)}, 
  title={The 2nd Clarity Enhancement Challenge for Hearing Aid Speech Intelligibility Enhancement: Overview and Outcomes}, 
  year={2023},
  volume={},
  number={},
  pages={1-5},
  doi={10.1109/ICASSP49357.2023.10094918}}

@INPROCEEDINGS{SI_of_SE_algorithms,
  author={Hu, Yi and Loizou, Philipos C.},
  booktitle={2007 IEEE International Conference on Acoustics, Speech and Signal Processing - ICASSP '07}, 
  title={A Comparative Intelligibility Study of Speech Enhancement Algorithms}, 
  year={2007},
  volume={4},
  number={},
  pages={IV-561-IV-564},
  doi={10.1109/ICASSP.2007.366974}}

@misc{ITUTP800,
    title = "{ITU-T Recommendation P.800: Methods for subjective determination of transmission quality}",
  author = "{International Telecommunication Union}",
    year = {1996},
month = {August}
}

@inproceedings{
ragano2024scoreq,
title={{SCOREQ}: Speech Quality Assessment with Contrastive Regression},
author={Alessandro Ragano and Jan Skoglund and Andrew Hines},
booktitle={The Thirty-eighth Annual Conference on Neural Information Processing Systems},
year={2024},
url={https://openreview.net/forum?id=HDVsiUHQ1w}
}

@article{schmidt1995intelligibility,
  title={Intelligibility and acceptability testing for speech technology},
  author={Schmidt-Nielsen, Astrid},
  journal={Applied speech technology},
  pages={194--231},
  year={1995},
  publisher={CRC press Boca Raton}
}

@misc{ANSIASAS3.2:2020_drt_mrt_pbwt,
  key = {ANSI/ASA S3.2-2020},
  title = {{ANSI/ASA S3.2-2020}: Method for Measuring the Intelligibility of Speech over Communication Systems},
  shorttitle = {ANSI/ASA S3.2-2020},
  author = {{Acoustical Society of America}},
  year = {2020},
  institution = {American National Standards Institute},
}

@article{zeghidour2022_soundstream,
  author={Zeghidour, Neil and Luebs, Alejandro and Omran, Ahmed and Skoglund, Jan and Tagliasacchi, Marco},
  journal={IEEE/ACM Transactions on Audio, Speech, and Language Processing}, 
  title={SoundStream: An End-to-End Neural Audio Codec}, 
  year={2022},
  volume={30},
  number={},
  pages={495-507},
  doi={10.1109/TASLP.2021.3129994}}

@inproceedings{barker25_clarity,
  title     = {{The 3rd Clarity Prediction Challenge: A machine learning challenge for hearing aid intelligibility prediction}},
  author    = {Jon Barker and Michael A. Akeroyd and Trevor J. Cox and John F. Culling and Jennifer Firth and Simone Graetzer and Graham Naylor},
  year      = {2025},
  booktitle = {{The 6th Clarity Workshop on Improving Speech-in-Noise for Hearing Devices (Clarity-2025)}},
}

@inproceedings{valin2019lpcnet,
  title={{LPCNet}: Improving neural speech synthesis through linear prediction},
  author={Valin, Jean-Marc and Skoglund, Jan},
  booktitle={ICASSP 2019 - 2019 IEEE International Conference on Acoustics, Speech and Signal Processing (ICASSP)},
  pages={5891--5895},
  year={2019},
  organization={IEEE}
}

@inproceedings{kumar23_dac,
 author = {Kumar, Rithesh and Seetharaman, Prem and Luebs, Alejandro and Kumar, Ishaan and Kumar, Kundan},
 booktitle = {Advances in Neural Information Processing Systems},
 pages = {27980--27993},
 publisher = {Curran Associates, Inc.},
 title = {High-Fidelity Audio Compression with Improved {RVQGAN}},
 volume = {36},
 year = {2023}
}

@inproceedings{shechtman24_dac_ibm,
  title     = {Low Bitrate High-Quality RVQGAN-based Discrete Speech Tokenizer},
  author    = {Slava Shechtman and Avihu Dekel},
  year      = {2024},
  booktitle = {Interspeech 2024},
  pages     = {4174--4178},
  doi       = {10.21437/Interspeech.2024-2366},
  issn      = {2958-1796},
}

@inproceedings{wu25f_ts3codec,
  title     = {{TS3-Codec: Transformer-Based Simple Streaming Single Codec}},
  author    = {Haibin Wu and Naoyuki Kanda and Sefik {Emre Eskimez} and Jinyu Li},
  year      = {2025},
  booktitle = {{Interspeech 2025}},
  pages     = {604--608},
  doi       = {10.21437/Interspeech.2025-921},
  issn      = {2958-1796},
}

@inproceedings{parker2025_stablecodec,
title={Scaling Transformers for Low-Bitrate High-Quality Speech Coding},
author={Julian D Parker and Anton Smirnov and Jordi Pons and CJ Carr and Zack Zukowski and Zach Evans and Xubo Liu},
booktitle={The Thirteenth International Conference on Learning Representations},
year={2025},
url={https://openreview.net/forum?id=4YpMrGfldX}
}

@ARTICLE{amr-wb,
  author={Bessette, B. and Salami, R. and Lefebvre, R. and Jelinek, M. and Rotola-Pukkila, J. and Vainio, J. and Mikkola, H. and Jarvinen, K.},
  journal={IEEE Transactions on Speech and Audio Processing}, 
  title={The adaptive multirate wideband speech codec ({AMR-WB})}, 
  year={2002},
  volume={10},
  number={8},
  pages={620-636},
  doi={10.1109/TSA.2002.804299}}

@INPROCEEDINGS{evs,
  author={Bruhn, S. and Pobloth, H. and Schnell, M. and Grill, B. and Gibbs, J. and Miao, L. and Järvinen, K. and Laaksonen, L. and Harada, N. and Naka, N. and Ragot, S. and Proust, S. and Sanda, T. and Varga, I. and Greer, C. and Jelínek, M. and Xie, M. and Usai, P.},
  booktitle={2015 IEEE International Conference on Acoustics, Speech and Signal Processing (ICASSP)}, 
  title={Standardization of the new {3GPP} {EVS} codec}, 
  year={2015},
  volume={},
  number={},
  pages={5703-5707},
  doi={10.1109/ICASSP.2015.7179064}}

@ARTICLE{liu24_semanticodec,
  author={Liu, Haohe and Xu, Xuenan and Yuan, Yi and Wu, Mengyue and Wang, Wenwu and Plumbley, Mark D.},
  journal={IEEE Journal of Selected Topics in Signal Processing}, 
  title={SemantiCodec: An Ultra Low Bitrate Semantic Audio Codec for General Sound}, 
  year={2024},
  volume={18},
  number={8},
  pages={1448-1461},
  doi={10.1109/JSTSP.2024.3506286}
}

@misc{defossez2024_moshi,
      title={Moshi: A speech-text foundation model for real-time dialogue}, 
      author={Alexandre Défossez and Laurent Mazaré and Manu Orsini and Amélie Royer and Patrick Pérez and Hervé Jégou and Edouard Grave and Neil Zeghidour},
      year={2024},
}

@INPROCEEDINGS{lyra,
  author={Kleijn, W. Bastiaan and Storus, Andrew and Chinen, Michael and Denton, Tom and Lim, Felicia S. C. and Luebs, Alejandro and Skoglund, Jan and Yeh, Hengchin},
  booktitle={ICASSP 2021 - 2021 IEEE International Conference on Acoustics, Speech and Signal Processing (ICASSP)}, 
  title={Generative Speech Coding with Predictive Variance Regularization}, 
  year={2021},
  volume={},
  number={},
  pages={6478-6482},
  doi={10.1109/ICASSP39728.2021.9415120}}

@misc{ITU-BS1534-3-2015,
  author = "{International Telecommunication Union}",
  title = "{ITU-R Recommendation BS.1534-3: Method for the subjective assessment of intermediate quality level of audio systems}",
  year = {2015},
  month = {October},
}

@misc{ITU-T_P.1401,
  author = "{International Telecommunication Union}",
  title = "{ITU-T Recommendation P.1401: Methods, metrics and procedures for statistical evaluation, qualification and comparison of objective quality prediction models}",
  number = {P.1401},
  year = {2020},
  month = {January},
}

@inproceedings{chinen2020visqol,
  title={{ViSQOL} v3: An open source production ready objective speech and audio metric},
  author={Chinen, Michael and Lim, Felicia SC and Skoglund, Jan and Gureev, Nikita and O'Gorman, Feargus and Hines, Andrew},
  booktitle={2020 twelfth international conference on quality of multimedia experience (QoMEX)},
  pages={1--6},
  year={2020},
  organization={IEEE}
}

@article{jensen2016_estoi,
  title={An Algorithm for Predicting the Intelligibility of Speech Masked by Modulated Noise Maskers}, 
  author={Jensen, Jesper and Taal, Cees H},
  journal={IEEE/ACM Transactions on Audio, Speech, and Language Processing},
  volume={24},
  number={11},
  pages={2009--2022},
  year={2016},
  publisher={IEEE}
}

@article{Graetzer2022_clarity,
  title = {{Dataset of British English speech recordings for psychoacoustics and speech processing research: The clarity speech corpus}},
  author = {Graetzer, Simone and Akeroyd, Michael A. and Barker, Jon and Cox, Trevor J. and Culling, John F. and Naylor, Graham J. and Pancani, Eszter and Tu, Zehai and Viveros-Muñoz, Raúl and Whitmer, William M.},
  journal = {Data in Brief},
  volume = {41},
  pages = {107951},
  year = {2022},
  month = {Feb},
  publisher = {Elsevier},
  doi = {10.1016/j.dib.2022.107951},
}

@misc{sekoyan2025_canary1bv2_parakeettdt06bv3,
      title={Canary-1B-v2 \& Parakeet-TDT-0.6B-v3: Efficient and High-Performance Models for Multilingual ASR and AST}, 
      author={Monica Sekoyan and Nithin Rao Koluguri and Nune Tadevosyan and Piotr Zelasko and Travis Bartley and Nikolay Karpov and Jagadeesh Balam and Boris Ginsburg},
      year={2025},
      eprint={2509.14128},
      archivePrefix={arXiv},
      primaryClass={cs.CL},
      url={https://arxiv.org/abs/2509.14128}, 
}

@manual{ITU_G191_2005,
  title        = {Recommendation G.191: Software Tools for Speech and Audio Coding Standardization},
  author       = {{International Telecommunication Union}},
  organization = {ITU-T},
  address      = {Geneva, Switzerland},
  month        = {November},
  year         = {2005},
}

@inproceedings{thiemann_demand,
  TITLE = {{The Diverse Environments Multi-channel Acoustic Noise Database (DEMAND): A database of multichannel environmental noise recordings}},
  AUTHOR = {Thiemann, Joachim and Ito, Nobutaka and Vincent, Emmanuel},
  BOOKTITLE = {{21st International Congress on Acoustics}},
  ADDRESS = {Montreal, Canada},
  ORGANIZATION = {{Acoustical Society of America}},
  YEAR = {2013},
  MONTH = {June},
  DOI = {10.5281/zenodo.1227120},
  HAL_ID = {hal-00796707},
  HAL_VERSION = {v1},
}

@inproceedings{schröter2022deepfilternet2,
  author={Schröter, H. and Maier, A. and Escalante-B, A.N. and Rosenkranz, T.},
  booktitle={2022 International Workshop on Acoustic Signal Enhancement (IWAENC)}, 
  title={Deepfilternet2: Towards Real-Time Speech Enhancement on Embedded Devices for Full-Band Audio}, 
  year={2022},
  volume={},
  number={},
  pages={1-5},
  doi={10.1109/IWAENC53105.2022.9914782}
}

@INPROCEEDINGS{suyanto2006_mltm,
  author={Suyanto},
  booktitle={TENCON 2006 - 2006 IEEE Region 10 Conference}, 
  title={Modified Least-to-Most Greedy Algorithm to Search a Minimum Sentence Set}, 
  year={2006},
  volume={},
  number={},
  pages={1-3},
  doi={10.1109/TENCON.2006.343825}}

@ARTICLE{ochiai2024_impact_SE_WER,
  author={Ochiai, Tsubasa and Iwamoto, Kazuma and Delcroix, Marc and Ikeshita, Rintaro and Sato, Hiroshi and Araki, Shoko and Katagiri, Shigeru},
  journal={IEEE/ACM Transactions on Audio, Speech, and Language Processing}, 
  title={Rethinking Processing Distortions: Disentangling the Impact of Speech Enhancement Errors on Speech Recognition Performance}, 
  year={2024},
  volume={32},
  number={},
  pages={3589-3602},
  doi={10.1109/TASLP.2024.3426924}}

@inproceedings{radford2023robust,
  title={Robust speech recognition via large-scale weak supervision},
  author={Radford, Alec and Kim, Jong Wook and Xu, Tao and Brockman, Greg and McLeavey, Christine and Sutskever, Ilya},
  booktitle={Proceedings of the 40th International Conference on Machine Learning},
 series = 	 {Proceedings of Machine Learning Research},
  pages={28492--28518},
  year={2023},
  organization={PMLR}
}

@INPROCEEDINGS{TaalSTOI,
  author={Taal, Cees H. and Hendriks, Richard C. and Heusdens, Richard and Jensen, Jesper},
  booktitle={2010 IEEE International Conference on Acoustics, Speech and Signal Processing}, 
  title={A short-time objective intelligibility measure for time-frequency weighted noisy speech}, 
  year={2010},
  volume={},
  number={},
  pages={4214-4217},
  doi={10.1109/ICASSP.2010.5495701}}

@article{nilsson1994development_hint,
  title={Development of the Hearing in Noise Test for the measurement of speech reception thresholds in quiet and in noise},
  author={Nilsson, Michael and Soli, Sigfrid D and Sullivan, Jean A},
  journal={The Journal of the Acoustical Society of America},
  volume={95},
  number={2},
  pages={1085--1099},
  year={1994},
  publisher={Acoustical Society of America},
  doi={10.1121/1.408469}
}

@article{billings2023speech_in_noise_testing,
  author = {Billings, Curtis J. and Olsen, Tina M. and Charney, Lenore and Madsen, Brandon M. and Holmes, Chloe E.},
  title = {Speech-in-Noise Testing: An Introduction for Audiologists},
  journal = {Seminars in Hearing},
  year = {2023},
  volume = {45},
  number = {1},
  pages = {55--82},
  month = {September},
  doi = {10.1055/s-0043-1770155},
}
\endgroup

\end{document}